\documentclass[aps,prd,floatfix,superscriptaddress,preprintnumbers,nofootinbib,twocolumn,longbibliography]{revtex4-1}
\pdfoutput=1

\usepackage{amsmath, amssymb, amsfonts, graphicx, mathrsfs, verbatim, float, slashed, bbm, color, xcolor, xspace, enumerate, url, bbding, multirow, outlines,upgreek}
\usepackage[english]{babel}
\usepackage[colorlinks=true,linkcolor=black,citecolor=galacticcenterbubblegum]{hyperref}
\definecolor{galacticcenterbubblegum}{rgb}{0.8,0, 0.8}

\begin{document}

\title{Calorimetric Dark Matter Detection With Galactic Center Gas Clouds}

\author{Amit Bhoonah}
\affiliation{The McDonald Institute and Department of Physics, Engineering Physics, and Astronomy, Queen's University, Kingston, Ontario, K7L 2S8, Canada}
\affiliation{ETH Zurich, Ramistrasse 101, 8092 Zurich, Switzerland}

\author{Joseph Bramante}
\affiliation{The McDonald Institute and Department of Physics, Engineering Physics, and Astronomy, Queen's University, Kingston, Ontario, K7L 2S8, Canada}
\affiliation{Perimeter Institute for Theoretical Physics, Waterloo, Ontario, N2L 2Y5, Canada}

\author{Fatemeh Elahi}
\affiliation{School of Particles and Accelerators, Institute for Research in Fundamental Sciences IPM, Tehran, Iran}

\author{Sarah Schon}
\affiliation{The McDonald Institute and Department of Physics, Engineering Physics, and Astronomy, Queen's University, Kingston, Ontario, K7L 2S8, Canada}
\affiliation{Perimeter Institute for Theoretical Physics, Waterloo, Ontario, N2L 2Y5, Canada}

\begin{abstract}
We demonstrate that dark matter heating of gas clouds hundreds of parsecs from the Milky Way Galactic center provides a powerful new test of dark matter interactions. To illustrate, we set a leading bound on nucleon scattering for 10-100 MeV mass dark matter. We also constrain millicharged dark matter models, including those proposed to match the recent EDGES 21 cm absorption anomaly. For Galactic center gas clouds, galactic fields' magnetic deflection of electromagnetically charged dark matter is mitigated, because the magnetic fields around the Galactic center are poloidal, as opposed to being aligned parallel to the Milky Way disk. We discuss prospects for detecting dark matter using a population of Galactic center gas clouds warmed by dark matter.
\end{abstract}

\maketitle



\section{Introduction}
\label{sec:intro}

The nature of dark matter remains a compelling mystery for physicists and astronomers. A detection of non-gravitational dark matter interactions would have a profound impact on our understanding of cosmology and the fundamental structure of the universe.  

Typical terrestrial searches for dark matter at underground laboratories are sensitive to relatively weak dark matter interactions with known particles. On the other hand, above-ground searches are required to find dark matter which interacts so strongly that it repeatedly scatters with nuclei and electrons during its passage through the Earth's atmosphere and crust. If it is moving too slowly after scattering with an overburden of atmosphere and crust, dark matter will not deposit enough kinetic energy in underground detectors to be identified, especially in the case of sub-GeV mass dark matter, which begins its voyage with less initial kinetic energy. (For the case of very heavy dark matter, which can scatter many times with the Earth and still trigger detectors deep underground, see Ref.~\cite{Bramante:2018qbc}.) As a consequence, there are a number of astrophysical \cite{Chivukula:1989cc,Starkman:1990nj,Qin:2001hh,Hook:2017vyc} and above-ground searches \cite{Rich:1987st,McGuire:1994pq,Bernabei:1999ui,Wandelt:2000ad,Albuquerque:2003ei,Zaharijas:2004jv,Mack:2007xj,Erickcek:2007jv,Kouvaris:2014lpa,Davis:2017noy,Dick:2017mgd,Mahdawi:2017cxz,Kavanagh:2017cru,Hooper:2018bfw,Emken:2018run} that have the best sensitivity to dark matter with relatively strong interactions.

This article demonstrates that dark matter heating of gas clouds, located hundreds of parsecs from the Milky Way Galactic center, can be used as a potent new method to seek out dark matter's coupling to known particles. In the Milky Way halo, dark matter heavier than a keV has a higher temperature than the coldest atomic gas clouds. Thus, for a given background dark matter density and velocity, these cold gas clouds can be used as very sensitive calorimetric detectors of dark matter's interactions. Hereafter, we discuss neutral gas cloud cooling, some recent observations of cold gas clouds hundreds of parsecs from the center of the Milky Way, and use these to bound dark matter-nucleon interactions along with millicharged dark matter. We conclude with prospects for detecting dark matter with a network of cold gas clouds.

\section{Galactic Center Gas Clouds and Dark Matter}
\label{sec:cgc}

Dark matter in the Milky Way galactic halo with velocity $v_{\rm x}$ has a temperature 
\begin{align}
T_{\rm x}  \sim m_{\rm x} v_{\rm x}^2 \simeq  10^4~{\rm K}~ \left(\frac{m_{\rm x}}{{\rm MeV}} \right) \left(\frac{v_{\rm x}}{10^{-3}} \right)^2, \nonumber
\end{align}
which often exceeds the temperature of neutral hydrogen gas clouds, which cool to temperatures as low as ten Kelvin. (Throughout this document we take $k_B = \hbar = c = 1$.) Atomic gas clouds with temperatures in the range $10-10^3$ K cool via collisional excitation and subsequent de-excitation of atomic fine structure transitions, where at low temperatures the fine structure transitions of oxygen, carbon, and iron account for most of the cooling \cite{Maio:2007yf}. In this study we will use the gas cloud cooling rates derived in \cite{DeRijcke:2013zha}; we have found that for cold gas with hydrogen number densities ranging from $n_{\rm H} \sim (0.1-10)~{\rm cm^{-3}}$ and temperatures ranging from $T_{\rm g} \sim (10-1000)~{\rm K}$, the volumetric cooling rate ($VCR$) is well approximated by the function
\begin{align}
VCR \simeq 2 &\times 10^{-22} ~{\rm \frac{GeV}{s~cm^3}}~\left(\frac{n_{\rm H}}{{\rm cm^{-3}}} \right)^2 \nonumber \\ &\times {\rm Exp} \left[-42.9~{\rm Log}\left[T_{\rm g}/{\rm K} \right]^{-1.4} \right],
\label{eq:approxvcr}
\end{align}
where the number density squared term arises from the rate for inter-atomic collisions in the gas cloud, which drive the cooling. We stress that this function is only valid for the temperatures and number densities indicated above.

Since inter-atomic collisions become less frequent at lower temperatures, the cooling rates of gas clouds \emph{decrease} over time. Therefore, observing a cold gas cloud sets an upper bound on heating from external sources,
\begin{align}
VCR \geq VSHR + VDHR,
\label{eq:heatingbound}
\end{align}
where we have indicated the volumetric heating rate from both dark matter ($VDHR$) and standard gas cloud heating sources ($VSHR$) such as cosmic rays and the dust grain photoelectric effect \cite{2009ARAA..47...27K}. In sections \ref{sec:dmn} and \ref{sec:mdm}, we will set bounds on dark matter interactions with gas cloud particles, making the assumption that, in order for the cloud to have cooled, the dark matter heating of the gas cloud must at least have been less than the cooling rate $VDHR \leq VCR$. Accounting for normal heating sources leads to stronger bounds on dark matter heating, $VDHR \leq VCR-VSHR$; this is left to future work.

\begin{figure}[t!]
  \includegraphics[width=.49\textwidth]{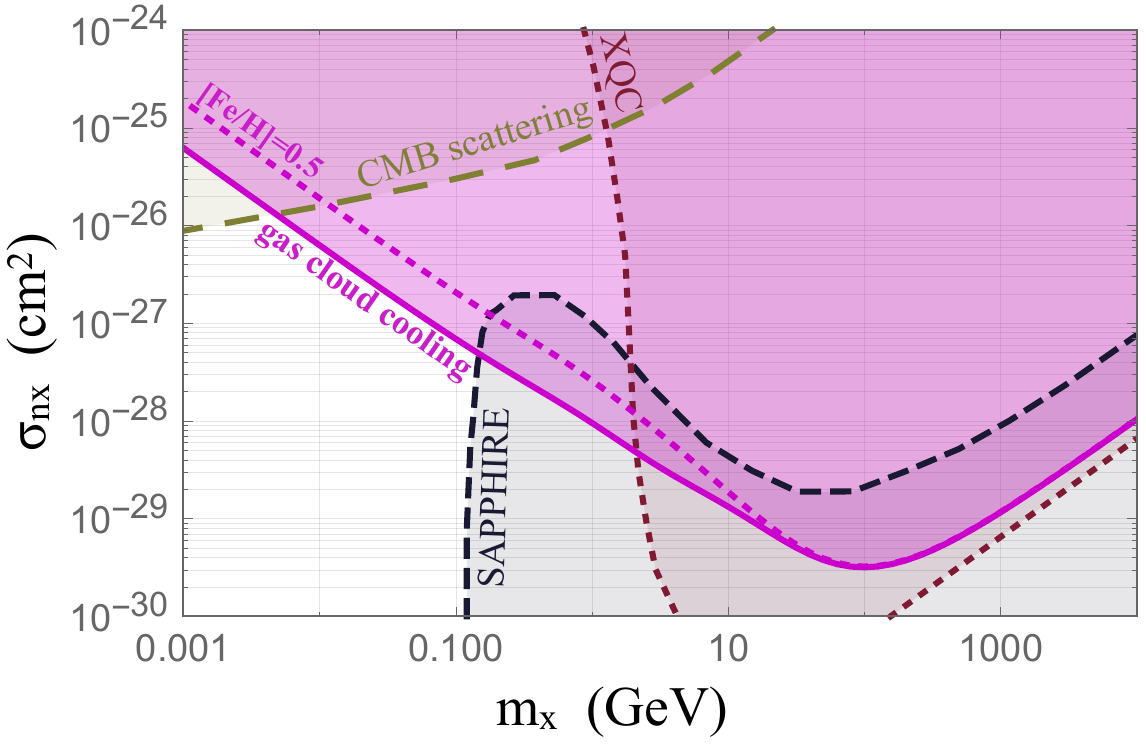}
  \caption{Sensitivity is shown for spin-independent dark matter scattering with nucleons in cold Galactic center gas clouds. Specifically, gas cloud G1.4-1.8+87 measured in Reference \cite{McClure-Griffiths:2013awa} sets the tightest constraint. We have shown the effect of increasing the gas cloud metallicity to the highest metallicity observed in a star, $[{\rm Fe/H}]=0.5$, although it is mostly likely the gas cloud will have solar or sub-solar metal content, resulting in a stronger bound, as discussed in the text. Alongside this bound, we plot representative bounds from dark matter detectors and cosmology. The dashed line labelled ``CMB scattering" indicates bounds from spectral distortions of the cosmic microwave background \cite{Dvorkin:2013cea,Gluscevic:2017ywp,Xu:2018efh,Slatyer:2018aqg}; we plot the results of Reference \cite{Gluscevic:2017ywp}. Terrestrial detection constraints from the above-ground run of CRESST-SAPPHIRE \cite{Davis:2017noy,Kavanagh:2017cru,Angloher:2017sxg} and the XQC Rocket \cite{Erickcek:2007jv,Mahdawi:2017cxz} are also indicated. For XQC we plot the 2\% XQC efficiency curve recently discussed in \cite{Mahdawi:2018euy}. For the sake of clarity, we omitted some overlapping bounds for $m_{\rm x} \gtrsim {\rm GeV}$, including DAMIC \cite{Barreto:2011zu} and the RRS balloon \cite{Rich:1987st}.}
  \label{fig:gcgcsi}
\end{figure}

Because for $\lesssim 10^3$ K gas clouds, cooling occurs mainly via fine structure transitions of oxygen, carbon, and iron, the atomic composition of cold gas clouds will affect the cooling rate. Specifically, the cooling rate will scale linearly with metallicity, $VCR \propto 10^{[{\rm Fe/H} ]}$, where metallicity is defined as the logarithm of the iron to hydrogen number density ratio, $[{\rm Fe/H} ] \equiv {\rm Log_{10}} (n_{\rm Fe}/n_{\rm H})$, normalized to the metallicity of the sun, $[{\rm Fe/H} ]_{\rm sun} \equiv 0$. The cooling rate in Equation~\eqref{eq:approxvcr} and the bounds shown in Sections \ref{sec:dmn} and \ref{sec:mdm} assume a solar metallicity. Note that gas clouds in the Milky Way typically have a solar metallicity \cite{2013ApJ...777...19H}, and that rare, higher-than-solar metallicity stars (the highest observed have $[{\rm Fe/H} ] \lesssim 0.5$ \cite{2018ApJ...855L...5D}) have only been found in the central few parsecs of the Milky Way and in globular clusters. On the other hand, the gas clouds we consider reside $z > 100$ pc above the plane of the Milky Way, where to the contrary, lower-than-solar metallicity gas clouds have been observed \cite{2013ApJ...777...19H,1999Natur.402..388W}. While it is plausible that follow-on observations of the Galactic center gas clouds we discuss in this article will lead to different bounds on light dark matter scattering with nuclei, we find in Sections \ref{sec:dmn} and \ref{sec:mdm} that adjusting gas cloud metallicity does not change bounds on high mass dark matter nucleon scattering or millicharged dark matter, because in these cases increasing the metallicity equally increases dark matter heating and gas cloud cooling.

Recently, a population of gas clouds near the galactic were discovered using the Australia Telescope Compact Array and Green Banks Telescope \cite{McClure-Griffiths:2013awa,2018ApJ...855...33D}. Using extensive observation of their 21 cm emission, these 174 gas clouds have been observed with masses ranging from $\sim 1 - 10^{5}$ solar masses, cloud radii spanning $\sim 3-30$ parsecs, and temperatures from $\sim 20-10^5$ K. Initial \cite{McClure-Griffiths:2013awa} and follow-on  observations \cite{2018ApJ...855...33D} revealed a population of neutral hydrogen gas clouds with a kinematic distribution consistent with clouds entrained in a hot wind produced by star formation in the central molecular zone (inner $\sim 100$ pc) of the Milky Way. Stringent bounds on dark matter interactions with gas clouds will arise from the coldest, least dense gas clouds that are closest to the Galactic center, where dark matter is more abundant. We have found that gas clouds G1.4-1.8+87 and G359.9+2.5+95 are most sensitive to dark matter scattering; G1.4-1.8+87 is six times colder and so provides $\sim$200 times greater cross section sensitivity.  The mass $M$, gas cloud radius $r_{\rm gc}$, average nucleon number density $n_{\rm n}$, temperature $T_{\rm g}$, line-of-sight distance from the Galactic center $r_{\rm los}$, and local standard of rest velocity $v_0$, of the most sensitive cloud which we use in setting bounds, \mbox{G1.4-1.8+87}, are $M=311 {\rm ~M_{\odot}},$ $r_{\rm gc}=12~{\rm pc}$, $n_{\rm n}=0.3~{\rm cm^{-3}},$ $T_{\rm g} \lesssim 22~{\rm K},$ $r_{\rm los}=332 ~{\rm pc},$ $v_{\rm g}=87.2~{\rm km/s}$.

\section{Dark Matter-Nucleon Scattering}
\label{sec:dmn}

Dark matter that interacts with nucleons will collide against and heat atomic nuclei in gas clouds. Scattered atoms have a mean free path that is small compared to the gas cloud \cite{2007adc..book.....I} and so will thermalize efficiently. For dark matter particles scattering elastically with nuclei in clouds, the per-nucleus heating rate is
\begin{align}
DHR \approx n_{\rm x} \sigma_{\rm Nx} v_{\rm x} E_{\rm nr},
\label{eq:dmheat}
\end{align}
where $n_{\rm x} = \rho_{\rm x}/m_{\rm x}$ is the local number density of dark matter, $E_{\rm nr} \approx \mu_{\rm Nx}^2 v_{\rm x}^2/m_{\rm N}$ is the average energy transferred per elastic scattering interaction with dark matter of mass $m_{\rm x}$, nuclei of mass $m_{\rm N}$, and dark matter-nuclear reduced mass $\mu_{Nx}$; $\sigma_{\rm Nx}$ is the dark matter-nuclear cross-section. The customary spin-independent per-nucleon cross-section $\sigma_{\rm nx}$ is defined in terms of $\sigma_{\rm Nx}$ as \cite{Lewin:1995rx}
\begin{align}
\sigma_{\rm N x} =  A^2 F_{\rm A}^2(E_{\rm nr}) \frac{\mu_{\rm Nx}^2}{\mu_{\rm nx}^2} \sigma_{\rm nx},
\label{eq:form}
\end{align}
where $A$ is the number of nucleons per nucleus, $\mu_{\rm nx}$ is the dark matter-nucleon reduced mass, and the Helm form factor is
\begin{align}
F_{\rm A}^2(E_{\rm nr}) = \left(\frac{3 j_1(qr)}{qr} \right)^2 e^{-s^2q^2},
\end{align}
where $j_1$ is the Bessel function of the first kind, \mbox{$q = \sqrt{2 m_{\rm N} E_{\rm nr}}$}, \mbox{$r=\sqrt{r_{\rm n}^2-5s^2}$}, the nuclear size is \mbox{$r_{\rm n} = 1.2~{A^{1/3}}~{\rm fm}$}, and the nuclear skin depth is \mbox{$s \sim 1~{\rm fm}$}. 

For bounds on dark matter scattering presented in Figure \ref{fig:gcgcsi}, we integrate over Boltzmann distributed velocities, accounting for the predicted dark matter velocity at the cloud's location, $v_0$, and the cloud's velocity relative to the galactic rest frame, $v_{\rm g}$. We consider scattering off of the most abundant atomic elements in the gas cloud, hydrogen, helium, oxygen, and iron, with mass abundance fractions following solar metallicity, $f_{\rm A} = \{ {\rm f_H, f_{He},f_{O},f_{Fe}} \}=\{ {\rm 0.73, 0.26, 0.0097, 0.0014} \}$. Altogether, the bound on per-nucleon scattering presented in Figure \ref{fig:gcgcsi} is
\begin{align}
\sigma_{\rm nx} < &\frac{VCR }{n_{\rm n}}   \left[  \sum_A  \frac{f_{A} A^2 \mu_{\rm Nx}^4 n_{\rm x}}{\mu_{\rm nx}^2 m_{\rm N}} \int dv_{\rm x}^3 v_{\rm x}^3 F_{\rm A}^2 B (v_{\rm x},v_{\rm esc},y) \right]^{-1},
\end{align} 
where we sum over atomic elements $A$, $F_{\rm A}$ is the Helm form factor given above. The integral over velocities runs from zero to the dark matter escape velocity $v_{\rm esc} \sim 0.002$, $y \equiv {\rm cos }~ \theta$ indicates the angle between the dark matter wind and the gas cloud, and the Boltzmann distribution is \cite{Lewin:1995rx}
\begin{align}
B= \frac{{\rm Exp} [-\left( v_{\rm x}^2 + v_{\rm g}^2 + 2v_{\rm x}v_{\rm g} y \right)]}
{\pi^{3/2} v_{\rm 0}^3 \left({\rm Erf}\left[\frac{v_{\rm esc}}{v_{\rm 0}} \right] - \frac{2v_{\rm esc}}{\pi^{1/2}v_{\rm 0}} {\rm Exp} \left[ \frac{-v_{\rm esc}^2}{v_{\rm 0}^2} \right] \right)}
\label{eq:boltz}
\end{align}
where this is normalized so that $\int dv_{\rm x}^3 B = 1$, the dark matter velocity dispersion is $v_{\rm 0} \approx 6\times 10^{-4}$ hundreds of parsecs from the Milky Way Galactic center \cite{Sofue:2013kja}, and the gas cloud-Milky Way relative velocity is $v_{\rm g} \sim 90~{\rm km/s}$ for gas cloud \mbox{G1.4-1.8+87} \cite{McClure-Griffiths:2013awa}. 

Figure \ref{fig:gcgcsi} shows the effect of increasing the gas cloud metallicity to $[{\rm Fe/H}] =0.5$. For higher mass dark matter there is no effect on the bound, since heating occurs mainly via scattering with iron; heavy nuclei are heated $A^5$ times faster than hydrogen by high mass dark matter, $c.f.$ Eqs.~\eqref{eq:dmheat} and \eqref{eq:form}. Therefore, increasing cloud metallicity equally increases the cloud cooling rate and the dark matter heating rate, resulting in the same bound.

Dark matter heating of Galactic center gas clouds will depend on the background dark matter density. However, because it is located $\sim 400$ pc from the Galactic center, the background dark matter density for gas cloud G1.4-1.8+87 does not depend much on the choice of dark matter density profile. Using an NFW \cite{Navarro:1995iw} profile $\rho_{\rm x} = \rho_0/((r/r_0)(1+r/r_0)^2)$ with $\rho_0 = 0.28 ~{\rm GeV/cm^3} $ and with scale radius $r_0 =20$ kpc, the density of dark matter at radius $r_{\rm tot} \lesssim \sqrt{2}~ r_{\rm los} \simeq 450$ pc is $\rho_{\rm x} \sim 13 {\rm~GeV/cm^3}$. (We have taken the gas cloud to be a factor of $\sqrt{2}$ further from the Galactic center than its angular distance \cite{McClure-Griffiths:2013awa} to account for possible line-of-sight projection.) This can be compared to a Burkert \cite{Burkert:1995yz} profile, $\rho_{\rm x} = \rho_{\rm b}/((r/r_{\rm b})(1+r^2/r^2_{\rm b}))$ with an $r_{\rm b} = 3$ kpc core and density normalization $\rho_{\rm b} = 14 ~{\rm GeV/cm^3}$, which implies a background dark matter density of $\rho_{\rm x} \sim 12 ~{\rm GeV/cm^3}$. We see that the assumption of a very cored dark matter halo distribution will have a small effect on the bound.

\section{Millicharged Dark Matter}
\label{sec:mdm}

\begin{figure}[t!]
  \includegraphics[width=.49\textwidth]{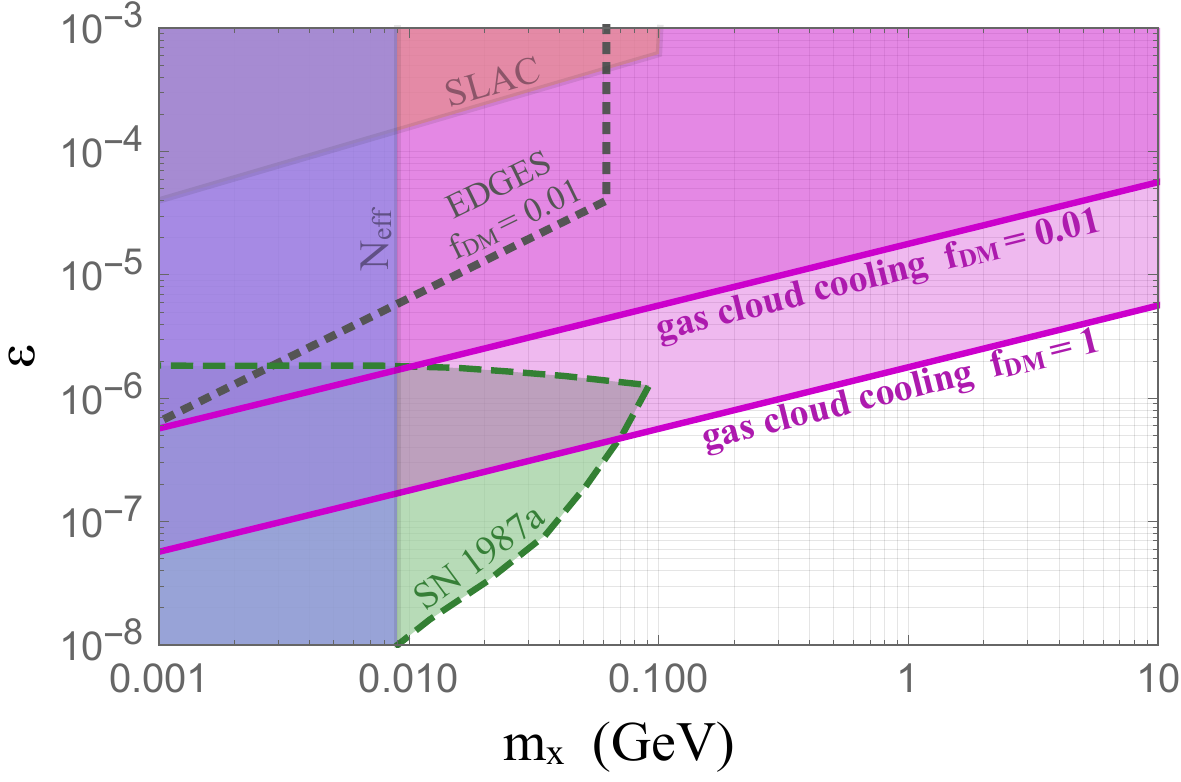}
  \caption{Constraints are shown on millicharged dark matter scattering with cold Galactic center gas clouds, for dark matter particles with charge $\epsilon \equiv Q/e$ which compose all ($f_{\rm DM}=1$) or one percent ($f_{\rm DM}=0.01$) of the dark matter. The metallicity of the gas cloud does not substantially alter the result, as explained in Section \ref{sec:mdm}. Galactic center gas cloud bounds on millicharged particles are likely unaffected by galactic magnetic fields \cite{Chuzhoy:2008zy,McDermott:2010pa}, because magnetic fields in the Galactic center are oriented towards the edge of the halo \cite{2010ApJ...722L..23N,2011ApJ...731...36L}, as opposed to being oriented parallel to the disk of the Milky Way. Parameter space which would explain the EDGES 21 cm anomaly for $f_{\rm DM}=0.01$ is shown with a dotted grey line, taken from \cite{Munoz:2018pzp} (see also \cite{Bowman:2018yin,Barkana:2018lgd,Munoz:2018pzp,Berlin:2018sjs,Barkana:2018qrx,Liu:2018uzy,Fraser:2018acy}). Bounds from supernova 1987a production of millicharged particles \cite{Chang:2018rso}, from the SLAC millicharged particle search \cite{Prinz:1998ua}, and from the millicharged particle contribution to the number of relativistic degrees of freedom in the early universe ($N_{\rm eff}$) \cite{Boehm:2013jpa} are indicated. For some recently reported experimental results on millicharged dark matter, see \cite{Crisler:2018gci,Magill:2018tbb,Mahdawi:2018euy}. For additional cosmological bounds, which can be sensitive to the millicharged fraction of dark matter ($f_{\rm DM}$), see \cite{Dubovsky:2003yn,McDermott:2010pa,Dvorkin:2013cea,Vogel:2013raa}.}
  \label{fig:gcgceps}
\end{figure}

Dark matter which effectively carries a small electromagnetic charge could arise from models where a new U(1) gauge boson mixes with the photon of the Standard Model \cite{Holdom:1985ag,Pospelov:2007mp,Pospelov:2008zw,Ackerman:mha,ArkaniHamed:2008qn,McDermott:2010pa,Davidson:2000hf}. Recently, millicharged dark matter has been proposed as an explanation of EDGES 21 cm data \cite{Bowman:2018yin,Barkana:2018lgd,Munoz:2018pzp,Berlin:2018sjs,Barkana:2018qrx,Liu:2018uzy,Fraser:2018acy}. Millicharged dark matter has been sought in astrophysical observations \cite{Dubovsky:2003yn,Chuzhoy:2008zy,McDermott:2010pa,Dvorkin:2013cea,Vogel:2013raa,Chang:2018rso} and experimental searches \cite{Prinz:1998ua,Crisler:2018gci,Magill:2018tbb,Mahdawi:2018euy}.

A particle with electromagnetic charge $\epsilon \equiv Q/e$ transiting a cold gas cloud will transfer energy to the gas cloud through a number of processes, chiefly by scattering with free electrons. The number of free electrons in a gas cloud with temperature $T_{\rm g} \lesssim 100$ K will depend on the abundance of carbon \cite{DeRijcke:2013zha}, which at low temperatures and for gas cloud densities $n_{\rm n} \sim 10^{-4}-100~{\rm cm^{-3}}$ will be either doubly or singly ionized \cite{DeRijcke:2013zha,1972ARA&A..10..375D,1995ApJ...443..152W} by the cosmic ultraviolet background \cite{FaucherGiguere:2009rh}. We will conservatively assume that the carbon atoms providing free electrons for scattering are singly-ionized -- for the temperature and densities considered here, a portion will actually be doubly-ionized \cite{DeRijcke:2013zha}. Then the number density of electrons in the gas cloud is the product of the nucleon number density and the carbon number density fraction, $n_{\rm e} = n_{\rm n} f_{\rm C} m_{\rm n} / m_{\rm C} \sim 6 \times 10^{-4} \times n_{\rm n}$. The Bethe formula \cite{Bethe:1930ku} gives the resulting energy transfer to the cloud,
\begin{align}
\frac{dE}{dx} = \frac{2 \pi n_{\rm e} \epsilon^2 \alpha_{\rm em}^2 }{m_{\rm e} v_{\rm x}^2} \left( {\rm~ ln}\left[ \frac{ 2 \mu_{\rm e x}^2 v_{\rm x}^2 / m_{\rm e}}{\lambda_{d}^{-2}/(2m_{\rm e})} \right]-v_{\rm x}^2 \right),
\end{align}
where $\alpha_{\rm em}$ is the fine structure constant, $n_{\rm e} ,m_{\rm e}$ are the number density and mass of electrons, $v_{\rm x}$ is the speed of the charged dark matter particle, and we have set the infrared cutoff of the Coulomb logarithm to account for plasma screening using the Debye length, \mbox{$\lambda_{\rm d} = \sqrt{T_{\rm g}/(4 \pi \alpha_{\rm em} n_{\rm e})} $}.
To derive a bound on millicharged dark matter heating a gas cloud with volume $V_{\rm c}$, we equate the total cooling rate of the gas cloud, $VCR \times V_{\rm c}$, with the heating expected from the total number of dark matter particles enclosed in the gas cloud, $n_{\rm x} V_{\rm c}$,
\begin{align}
VCR \approx \int dv_{\rm x}^3  \frac{2 \pi B (v_{\rm x}) n_{\rm x} n_{\rm e} \epsilon^2 \alpha_{\rm em}^2 }{m_{\rm e} v_{\rm x}} {\rm~ ln}\left[ \frac{ 2 \mu_{\rm e x}^2 v_{\rm x}^2 / m_{\rm e}}{\lambda_{d}^{-2}/(2m_{\rm e})} \right],
\end{align}
where we have divided both sides of this equation by the gas cloud volume, $V_{\rm c}$, and integrate over the Boltzmann distribution $B(v_{\rm x})$ given in Eq. \eqref{eq:boltz}.

The bound from gas cloud G1.4-1.8+87 on millicharged dark matter $\epsilon$ is displayed in Figure \ref{fig:gcgceps}. Note that varying gas cloud metallicity does not affect this result: both the gas cloud cooling rate and the free electron fraction depend linearly on the carbon content of the cloud \cite{DeRijcke:2013zha,1972ARA&A..10..375D,1995ApJ...443..152W}. Increasing cloud metallicity increases cloud cooling and millicharged dark matter heating by the same proportion. The carbon ionization fraction of the gas cloud is less debatable, since for the densities considered, clouds will be fully ionized at temperatures $T_{\rm g} \lesssim 100$ K by the ubiquitous ultraviolet background \cite{Maio:2007yf,FaucherGiguere:2009rh}. Nevertheless, the bound is also relatively insensitive to ionization fraction: both dark matter heating and atomic gas cooling increase with ionization \cite{DeRijcke:2013zha}.

It has been noted that $\sim 10~{\rm \mu Gauss}$ magnetic fields in the Milky Way Galactic disk, may deflect and eventually evacuate most millicharged dark matter from the Milky Way disk, for $\epsilon > 5 \times 10^{-13} (m_{\rm x} /{\rm GeV})$, since these magnetic fields are largely aligned parallel to the disk \cite{Chuzhoy:2008zy}. However, irregularities in these magnetic fields may still allow some dark matter to diffuse into the disk \cite{McDermott:2010pa}. At the Milky Way Galactic center, for galactic heights in excess of $z \gtrsim 100$ pc, it has been observed that, rather than being oriented parallel to the Milky Way halo, the magnetic fields have a poloidal orientation, $i.e.$ the fields point perpendicular to the Milky Way disk plane \cite{2010ApJ...722L..23N,2011ApJ...731...36L}. Hence, it appears that most millicharged dark matter from the Milky Way halo will flow un-deflected into the region surrounding Galactic center gas clouds.

\section{Discussion}

We have established Galactic center gas clouds as a puissant method for finding dark matter interactions with baryons and electrons. The resulting analysis has placed leading bounds on 10-100 MeV dark matter scattering with nucleons and on millicharged dark matter for a wide range of masses. 

In future work, it will be interesting to investigate how the remaining $\sim 173$ Galactic center gas clouds can be utilized to search for dark matter interactions. If dark matter has substantial enough interactions with cold gas clouds, this could become evident as a minimum gas cloud temperature, as a function of gas cloud density, and separation of the gas cloud from the Galactic center, where gas clouds at a greater distance from the Galactic center will be heated by less dark matter. Furthermore, the bounds and sensitivity of Galactic center gas clouds presented here can likely be improved, by more carefully determining the amount of heating from standard heating processes \cite{1995ApJ...443..152W,2007adc..book.....I}, like heating via photo-ejection of electrons from interstellar dust. It will also be interesting to investigate Galactic center gas cloud sensitivity to a variety of sub-GeV mass dark matter models \cite{Knapen:2017xzo,Berlin:2018tvf}.

\acknowledgments 

We especially wish to thank Judith Irwin and Kristine Spekkens for guidance on radio astronomy and HI clouds. We thank Prateek Agrawal, Michael Balogh, Asher Berlin, Kimberly Boddy, Michel Fich, Vera Gluscevic, Rafael Lang, Maxim Pospelov, Nirmal Raj, Tracy Slatyer, Yu-Dai Tsai, and Aaron Vincent for discussions. J.~B.~and F.~E.~thank the CERN theory group for hospitality. J.~B.~thanks the organizers of the KITP High Energy Physics at the Sensitivity Frontier Workshop; this research was supported in part by the NSF under Grant No.~PHY-1748958. Research at Perimeter Institute is supported by the Government of Canada through Industry Canada and by the Province of Ontario through the Ministry of Economic Development \& Innovation. A.~B., J.~B., and S.~S.~ acknowledge the support of the Natural Sciences and Engineering Research Council of Canada. J.~B.~thanks the Aspen Center for Physics, which is supported by NSF Grant No.~PHY-1066293.

\bibliography{gcgc}

\end{document}